\documentclass[twocolumn,showpacs,preprintnumbers,amsmath,amssymb,prb,aps]{revtex4-1}
\usepackage{epsfig}% Include figure files
\usepackage{epstopdf}
\usepackage{graphicx}% Include figure files
\usepackage{dcolumn}% Align table columns on decimal point
\usepackage{bm}% bold math
\usepackage{textcomp}
\usepackage{array}
\usepackage{subcaption}
\usepackage{booktabs}

\DeclareUnicodeCharacter{2212}{-}
\begin{document}

\title{Exploring temperature dependent electron-electron interaction of topological crystalline insulators (SnS and SnSe) within Matsubara-time domain}

\author{Antik Sihi$^{1,}$}
\altaffiliation{sihiantik10@gmail.com}
\author{Sudhir K. Pandey$^{2,}$}
\altaffiliation{sudhir@iitmandi.ac.in}
\affiliation{$^{1}$School of Basic Sciences, Indian Institute of Technology Mandi, Kamand - 175075, India\\
$^{2}$School of Engineering, Indian Institute of Technology Mandi, Kamand - 175075, India}

\date{\today}

\begin{abstract}

Both experimental and theoretical studies show non-trivial topological behaviour in native rocksalt phase for SnS and SnSe and categorize these materials in topological crystalline insulators. Here, the detailed electronic structures studies of SnS and SnSe in the rocksalt phase are carried out using many-body $GW$ based theory and density functional theory both for ground states and temperature dependent excited states. The estimated values of fundamental direct bandgaps around L-point using $G_0W_0$ (mBJ) are $\sim$0.27 ($\sim$0.13) eV and $\sim$0.37 ($\sim$0.17) eV for SnS and SnSe, respectively. The strength of hybridization between Sn 5$p$ and S 3$p$ (Se 4$p$) orbitals for SnS (SnSe) shows strong k-dependence. The behaviour of $\overline{W}$ ($\omega$), which is the averaged value of diagonal matrix elements of fully screened Coulomb interaction, suggests to use full-$GW$ method for exploring the excited states because the correlation effects within these two materials are relatively weak. The temperature dependent electronic structure calculations for SnS and SnSe provide linearly decreasing behaviour of bandgaps with rise in temperatures. The existence of collective excitation of quasiparticles in form of plasmon is predicted for these compounds, where the estimated values of plasmon frequency are $\sim$9.5 eV and $\sim$9.3 eV for SnS and SnSe, respectively. The imaginary part of self-energy and mass renormalization factor ($Z_\textbf{k}(\omega)$) due to electron-electron interaction (EEI) are also calculated along W-L-$\Gamma$ direction for both the materials, where the estimated ranges of $Z_\textbf{k}(\omega)$ are 0.70 to 0.79 and 0.71 to 0.78 for SnS and SnSe, respectively, along this k-direction. The present comparative study reveals that the behaviour of temperature dependent EEI for SnS and SnSe are the almost same and EEI is important for high temperature transport properties.    

\end{abstract}

\maketitle

\section{Introduction} 

\begin{figure*}
  \begin{center}
    \includegraphics[width=0.9\linewidth, height=9.5cm]{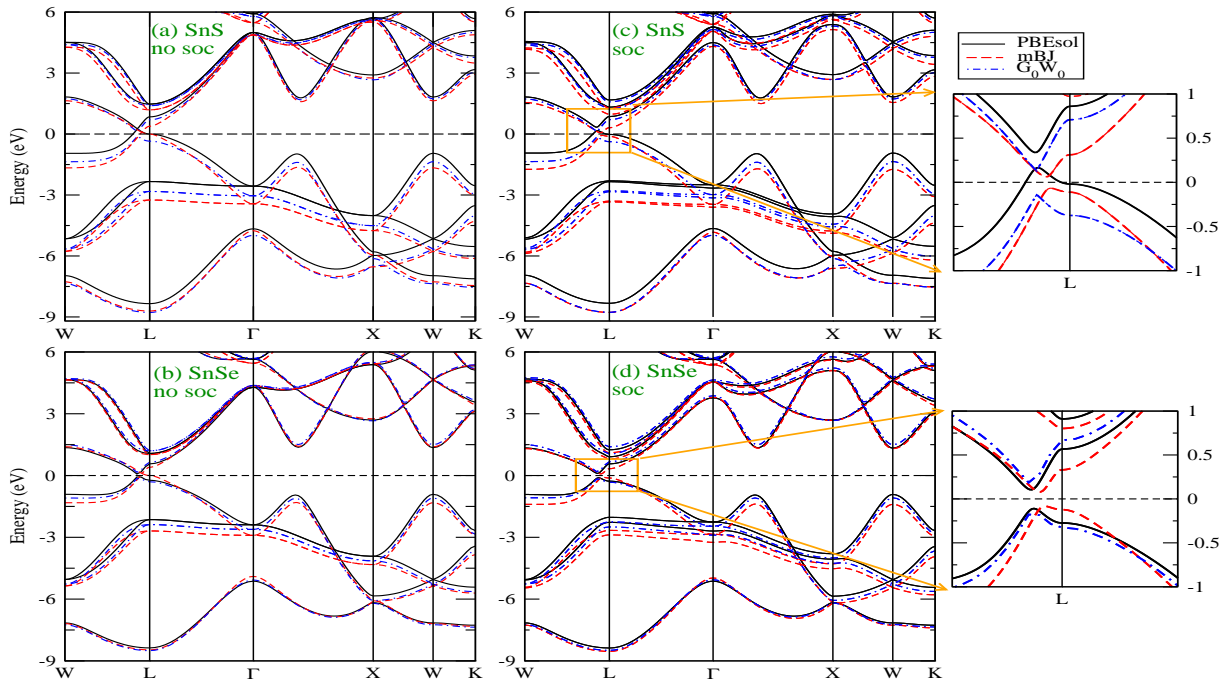} 
    \caption{(Colour online) Electronic band structures of SnS and SnSe (a) \& (b) without and (c) \& (d) with including spin-orbit coupling using PBEsol (black solid line), mBJ (red dashed line) and $G_0W_0$ (blue dotted dash line).}
    \label{fig:}
  \end{center}
\end{figure*}
    
 From past few decades, topological materials (TM) opened a new era of research in the field of condensed matter physics due to the presence of exotic phenomena and multiple application purposes they offer \cite{hasan_rev,bansil_rev}. Topological crystalline insulator (TCI) is one of the important members in the class of TM after mentioned by L. Fu. \cite{fuprl} TCI shows finite bandgap in the corresponding bulk band structure, whereas the conducting surface states of this material are usually protected by mirror symmetry. Tin chalcogenides (SnS, SnSe and SnTe) show non-trivial topological behaviour in native rocksalt structure without applying any external perturbation and belong to TCI family \cite{hsieh,tanaka,y_sun,wang_snse_exp1,jin_snse_exp2}. In light of this, the electronic structure of SnS and SnSe in rocksalt phase are not explored in details, whereas SnTe in the same phase has gained much attention. It is known that rocksalt SnS can be formed by the epitaxial growth after selecting proper substrate \cite{mariano_exp_sns, bilenkii_exp_sns2, skelton_sns}, and rocksalt SnSe is dynamically stable in ambient conditions \cite{wang_snse_exp1,jin_snse_exp2,wang_snse}. Besides presence of topological behaviour in this phase, SnSe is also theoretically predicted to be appropriate for using as optocaloric cooling device \cite{zhou_opto}. However as per our present knowledge, experimental evidence of bandgap values in rocksalt phase for SnS and SnSe are still not available for comparison. Therefore, proper theoretical estimations of bandgaps for these materials are important for enhancing the predictive power of various properties for practical applications. The reason of the bandgap problem of density functional theory (DFT) with local density approximation (LDA) or generalized gradient approximation (GGA) is known to electronic structure community. Thus, the methodology of mBJ potential within DFT formalism is developed for solving the bandgap issue \cite{koller2011,koller2012}. However, any DFT based method always provides the single-particle energies which normally does not properly estimate the quasiparticle energies of the interacting electron system. 
  
  The quasiparticle energies of interacting electrons can be computed through $GW$ approximation (GWA) based on many-body perturbation theory via calculating $\Sigma$ of the system. The $\Sigma$ reveals the information about the electron-electron interaction (EEI). Here, $G$ represents the one particle Green's function and $W$ denotes the fully screened Coulomb interaction. The one-shot GWA, which is known as $G_0W_0$, generally corrects the Kohn-Sham states by perturbative manner for obtaining the quasiparticle states of interacting electrons. The $G_0W_0$ method is typically famous for providing nice estimation of bandgap for different class of materials. To the best of our knowledge, the $G_0W_0$ corrected electronic structures for SnS and SnSe in topological crystalline insulating phase are not reported. However, the result obtained from the $G_0W_0$ method typically shows dependency on the choice of exchange-correlation (XC) functional at DFT level. Therefore at this point, fully self-consistent $GW$ (full-$GW$) method is expected to estimate the proper quasiparticle energies of interacting electrons, where the finally computed result doesn't depend on the starting point of single-particle Kohn-Sham states. Thus, exploring the bandgap problem along with studying the many-body interaction will provide the deeper understanding of any exotic phenomena and the applicability of any material. Moreover, investigation of temperature dependent electronic structure calculation always brings insightful information about the thermal effect on EEI of material.  
  
  Recently, few comparative works on different properties of SnS and SnSe are found from the literatures in rocksalt structure \cite{y_sun, pallikara, yakovkin}. The strength of spin-orbit coupling (SOC) along with the lattice parameters typically increase from SnS to SnSe and mainly responsible for observing the subtle difference for various properties between these two materials. Therefore in this context, it is interesting to investigate that if one moves form SnS to SnSe then whether the strength of correlation within the sample is also changing or not. The correlation strength along with the presence of plasmon excitation can be possibly estimated from the frequency ($\omega$) dependent $W$ \cite{antik, arya2006, miyake2009, sakuma2013, amadon2014}. Also, the comparative studies of many-body interactions using full-$GW$ method may illustrate more clear view of electronic structure's properties of these two materials.     
   
 In this work, we explored both, ground and excited electronic structures of SnS and SnSe in rocksalt phase using many-body theory based on $G_0W_0$ and full-$GW$ methods. The comparative study between DFT and $G_0W_0$ methods for ground state calculations is also carried out. The importance of SOC for estimating the values of bandgaps are seen for both the materials. The calculated value of bandgap for SnS using $G_0W_0$ (mBJ) is $\sim$0.13 ($\sim$0.27) eV, where the fundamental bandgap is not seen from PBEsol functional. In case of SnSe, the computed values of bandgaps is found to be $\sim$0.22 eV, $\sim$0.17 eV and $\sim$0.37 eV for PBEsol, mBJ and $G_0W_0$, respectively. The strength of hybridization between Sn 5$p$ and S 3$p$ (Se 4$p$) orbitals for SnS (SnSe) has shown strong dependency on the k-point. The correlation strength of SnS and SnSe is estimated from the $\omega$ dependent $\overline{W}$, which defines the averaged value of diagonal elements of fully screened Coulomb interaction. The different forms of Coulomb interactions are calculated using random phase approximation for both the materials. The temperature dependent bandgap along with spectral functions due to EEI are explained. The presence of plasmon excitations are predicted from the full-$GW$ calculations for SnS and SnSe. The detailed studies of many-body EEI for both TCIs are carried out by explaining the temperature and \textbf{k}-point ($i.e.$ along W-L-$\Gamma$ direction) dependent imaginary part of the self-energy ($Im \Sigma (\textbf{k},\omega)$). The similar amount of EEI is evident from the present study for both materials. Momentum-resolved spectral function is also discussed for SnS and SnSe.   
  
\section{Theoretical Methods}   

\subsection{Description of different Coulomb interactions}

  Random phase approximation (RPA) is used for estimating different Coulomb interactions. In order to calculate these quantities, we need to form the local atomic-like orbitals from the Kohn-Sham orbitals. Maximally localized Wannier function (MLWF) is chosen to build the local orbitals for particular bands which are found in certain energy region. The orbitals of dominating characters around the $E_F$ are usually taken for obtaining the MLWF. In general, the fully screened Coulomb interactions ($W$) and bare Coulomb interaction ($\upsilon$) are given below \cite{vaugier},
  
\begin{equation}
W^{\mathbf{R}_1\mathbf{R}_2\mathbf{R}_3\mathbf{R}_4}_{L_1L_2L_3L_4} (\omega) \equiv \langle \phi_{\mathbf{R}_1L_1}\phi_{\mathbf{R}_2L_2}|W(\omega)|\phi_{\mathbf{R}_3L_3}\phi_{\mathbf{R}_4L_4} \rangle 
\end{equation} 

\begin{equation}
\upsilon ^{\mathbf{R}_1\mathbf{R}_2\mathbf{R}_3\mathbf{R}_4}_{L_1L_2L_3L_4} \equiv \langle \phi_{\mathbf{R}_1L_1}\phi_{\mathbf{R}_2L_2}|\upsilon|\phi_{\mathbf{R}_3L_3}\phi_{\mathbf{R}_4L_4} \rangle 
\end{equation}      

 where $L$ = ($m,\alpha,\sigma$), $m$ is an orbital quantum number, $\alpha$ represents the atomic index of a given unit cell centered at $\mathbf{R}$ and spin degree of freedom is denoted by $\sigma$. $\phi_{\mathbf{R}L}$ represents the MLWF of corresponding Kohn-Sham orbital. The relation between $W(\omega)$ and $\upsilon$ is found to be $W = [1 - \upsilon P]^{-1} \upsilon$, where $P$ represents the total polarization within RPA and $\upsilon$ is the Coulomb interaction between two bare charges \cite{martin}. It is noted that the $W$ is a function of $\omega$, since $P$ depends on $\omega$. The on-site Coulomb interactions are obtained when we fixed $\mathbf{R}_1$=$\mathbf{R}_2$=$\mathbf{R}_3$=$\mathbf{R}_4$. Therefore in case of on-site intra-atomic, the notation $L$ only depends on the value of $m$ and the on-site bare exchange Coulomb interaction ($J_{bare}$) is denoted by, 

\begin{equation}
J_{bare} \equiv J_{bare}^{m,m'} = \langle \phi_{m}\phi_{m'}|W(0)|\phi_{m'}\phi_{m} \rangle 
\end{equation}

\subsection{Discussion on Matsubara-time domain self-consistent $GW$ method}

  Now, it is well understood that preforming self-consistent $GW$ calculation in Matsubara-time domain is more easy and computationally convenient to implement than the conventional statistical mechanics approach. On other hand, the effect of temperature on the electronic structure properties of any material is also possible to study using the former formalism \cite{landau}. This Matsubara-time Green's function method is normally utilized within DFT + DMFT method form few decades \cite{kotlier,pdjpcm,pdepl}. In continuation with this, the information about the many-body effect due to EEI can be obtained from the calculated values of electronic self-energy ($\Sigma$) using this methodology. The $\Sigma(\tau)$ is obtained from the self-consistent calculation, where $\tau$ denotes the Matsubara-time. Next, it is transferred into $\Sigma(i\omega_n)$ using Fourier transformation, where $\omega_n=(2n+1)\frac{\pi}{\beta}$ for fermions. Here, $n$ is integer, $\beta=\frac{1}{k_BT}$ and $T$ represents temperature. But, in order to obtain the spectral function in real $\omega$, the analytic continuation is generally performed to transform $\Sigma(i\omega_n)$ to $\Sigma(\omega+i\eta)$, where $\eta$ denotes a positive infinitesimal number.

\section{Computational details}

 Here, WIEN2k \cite{wien2k} code is chosen to carry out the spin-unpolarized electronic structure calculation for SnS and SnSe compounds. This code is based on full-potential linearized-augmented plane-wave (FP-LAPW). The space group of $F$m-3m is used in present study for both materials. The PBEsol XC functional \cite{pbesol} and mBJ exchange potential with GGA correlation \cite{mbj} are utilized for DFT calculation. The estimated values of optimized lattice parameters, which are computed using PBEsol functional for SnS and SnSe, are 5.753\AA \,and 5.955\AA \,, respectively. It is noted that these values are showing good agreement with the previously reported theoretical and experimental data of these two materials \cite{y_sun,mariano_exp_sns,bilenkii_exp_sns2,burton}. The self-consistent DFT calculations are performed on 10\texttimes 10\texttimes 10 \textbf{k}-mesh size. The Wyckoff positions for Sn and S (Se) are fixed at (0.0, 0.0 ,0.0) and (0.5, 0.5, 0.5), respectively, where the muffin-tin radius are set to be 2.5 Bohr for Sn and Se along with 2.43 Bohr for S. The convergence criteria is kept to 10$^{-4}$ Ry/cell for calculating the total energy. In order to get the better convergence, $R_{mt} * K_{max}$ is considered as 8.5 in whole calculations. GAP2 code based on Wannier basis function is utilized for computing different Coulomb interaction with the help of RPA \cite{jiang1, jiang2}. GAP2 is also used for carrying out $G_0W_0$ calculations on both the compounds. In addition to this, Elk \cite{elk} code is chosen for performing the full-$GW$ calculation, where the Matsubara-time domain is implemented to perform the calculation on imaginary axis. In order to get the spectral function on real axis, Pa$\acute{d}$e approximation \cite{pade} is employed to transform the data from imaginary to real axis. The procedure of analytic continuation based on Pa$\acute{d}$e approximation is chosen due to its' lower computational cost and simple implementation within the full-$GW$ methodology. 4\texttimes 4\texttimes 4 \textbf {q}-mesh size is taken together with 8\texttimes 8\texttimes 8 \textbf{k}-mesh for full-$GW$ calculation. It is known that the $G_0W_0$ calculation depends on using the starting XC functional at DFT level. Therefore, to get rid from this issue, we keep PBEsol XC functional at DFT level for all types of $GW$ based calculations \cite{pbesol}. Here, the ground state electronic structures of SnS and SnSe are studied both in presence and absence of SOC. 
  
\begin{figure}
  \begin{center}
    \includegraphics[width=0.85\linewidth, height=10.0cm]{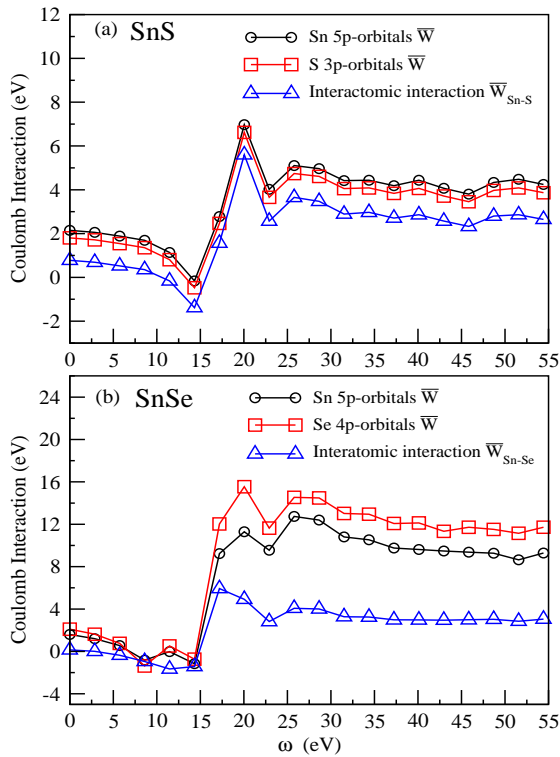} 
    \caption{(Colour online) Coulomb interactions as a function of $\omega$ for SnS and SnSe.}
    \label{fig:}
  \end{center}
\end{figure}

\section{Results and Discussion} 

\subsection{Ground state electronic structure and correlation effect}
 
  To understand the basic electronic structure of any non-trivial TCI, the detailed discussion on ground state properties with proper estimation of fundamental bandgap is known to be important. Therefore, the systematic theoretical studies on both SnS and SnSe are carried out in present work with different functionals at DFT level together with different methods ($i.e.$ DFT, $G_0W_0$). Here, it is noted that zero energy in the dispersion plot represents the Fermi energy ($E_F$). The $E_F$ is fixed at middle of the bandgap when a clear bandgap is seen from any calculated band structure. Figs. 1(a) and (b) show the dispersion curves for SnS and SnSe along the high symmetric k-direction, which are calculated using PBEsol, mBJ and $G_0W_0$ without considering SOC. In case of conduction band (CB), the bands of PBEsol and $G_0W_0$ are showing almost similar behaviour along the observed k-direction for both samples. For SnS in CB region, the shifting of mBJ bands toward the lower energy with respect to PBEsol bands has been clearly seen, where the maximum shifting is found around the L-point. But for SnSe, the mBJ bands nicely followed the PBEsol bands for CB region. Similarly for valence band (VB) region, the bands obtained from $G_0W_0$ move towards to lower energy than PBEsol for both materials, whereas the further shifting of the bands to lower energy is seen for mBJ bands with respect to $G_0W_0$. The energy difference between mBJ ($G_0W_0$) bands and PBEsol bands is estimated to be maximum for W-L-$\Gamma$ direction, where the calculated differences are 0.6 - 0.9 (0.4 - 0.5) eV and 0.4 - 0.53 (0.2 - 0.22) eV for SnS and SnSe, respectively. Here, it is noted that the energy shifting is more prominent for SnS than SnSe, which may be due to the increase of lattice parameters from SnS to SnSe. In addition to this for SnS, $G_0W_0$ shows direct bandgap of amount $\sim$0.23 eV near the L-point, whereas PBEsol and mBJ provide metallic behaviour. In case of SnSe, the calculated values of bandgap using PBEsol and $G_0W_0$ are found to be $\sim$0.15 eV and $\sim$0.32 eV, respectively and the band structure of mBJ shows metallic nature. Moreover, due to the presence of heavier element Sn in both the compounds, SOC is expected to play important role for predicting the better electronic ground state. Therefore, we have performed the further ground state calculation including SOC. Here, the SOC correction at DFT level is considered in $G_0W_0$ method by perturabtive manner. The calculated dispersion curves using PBEsol, mBJ and $G_0W_0$ including SOC are plotted in Figs. 1 (c) and (d) for SnS and SnSe, respectively. The band features are almost similar with the previous discussion ($i.e.$ without including SOC) except some band degeneracy lifting along the high-symmetric k-direction due to SOC. This behaviour is usually observed after including SOC. The PBEsol does not show any bandgap after including SOC for SnS, whereas mBJ ($G_0W_0$) creates bandgap of $\sim$0.13 ($\sim$0.27) eV around L-point. Similarly for SnSe after inclusion of SOC, it is seen that PBEsol, mBJ and $G_0W_0$ provide the bandgap of $\sim$0.22 eV, $\sim$0.17 eV and $\sim$0.37 eV, respectively. The fundamental bandgap of many-body electronic system is usually defined as the difference in calculated energy between adding an electron and removing an electron from the sample system \cite{martin}. Therefore, the bandgap obtained from PBEsol is not expected to provide correct value as compared to experimental data because the orbital dependency for computing the band energy at conduction band minima and valence band maxima is missing in Kohn-Sham implementation along with the problem of derivative discontinuity. But at least, it is desired that PBEsol will show semiconducting behaviour for both SnS and SnSe due to the presence of $p$-like electrons at the lower energy region around $E_F$. However, the dispersion plot of PBEsol of SnS shows metallic behaviour, but the semiconducting nature is seen for SnSe. Since 3$p$ orbitals are more localized than 4$p$ orbitals, therefore the correlation between the electrons of S 3$p$ orbitals might be improperly estimated for SnS compound. Thus, it leads to the metallic behaviour of SnS even after including SOC. Now, in case of mBJ, it shows the semiconducting behaviour for both the materials by correcting the kinetic energy density of these compounds. But, the mBJ method also comes under the Kohn-Sham formalism, where the electronic screening is not considered within the methodology. This screening is important for finding the bandgap of any material. Hence, it is also not expected from mBJ to give proper fundamental bandgap as compared to experimental bandgap \cite{koller2011, jiang_mbj}. At this point, it is important to note that the $G_0W_0$ based on many-body perturbation theory is known to provide better estimation of bandgap, where all the aforementioned drawbacks of Kohn-Sham formalism are overcome. Therefore, the estimated values of fundamental bandgaps using $G_0W_0$ for both samples will be expected to match with experimental data. It is also noted that SOC plays important role for estimating the bandgap for both these materials. In addition to this, the calculated values of direct bandgaps using $G_0W_0$ at L-point are found to be $\sim$1.08 eV and $\sim$0.98 eV for SnS and SnSe, respectively. The fundamental bandgap as computed using $G_0W_0$  method for SnS is showing smaller value than SnSe around the L-point, but the direct bandgap at L-point for former compound provides higher value than the other one. Thus, it is evident for L-point (the k-point where the fundamental bandgap is observed) that the hybridization strength between Sn 5$p$ and S 3$p$ orbitals for SnS is stronger (weaker) than Sn 5$p$ and Se 4$p$ orbitals for SnSe. At this point, it is important to note that the hybridization strength of these materials show dependency on the choice of k-point within first Brillouin zone. The 5$p$ and 3$p$ (4$p$) orbitals of Sn and S (Se) are mainly responsible for making the energy bands within the energy window of -8.5 eV (-6.0 eV) to 8.0 eV (7.0 eV) for SnS (SnSe) material. These energy bands around the $E_F$ are usually expected to participate for showing the transport and topological behaviours. Presence of electronic correlation always brings new exotic behaviour along with these mentioned properties. Therefore, investigation of different Coulomb interactions by considering these orbitals will provide insightful information about the correlation effects of both compounds. 

\begin{table}
\caption{\small{Averaged values of diagonal and off-diagonal matrix elements of bare Coulomb interaction ({$\overline{\upsilon}$}) for Sn (S) 5$p$ (3$p$) orbitals \& matrix elements of bare ({$\overline\upsilon_{Sn-S}$}) and screened ({$\overline W_{Sn-S}$}) inter-atomic Coulomb interactions. The on-site bare exchange interaction $J_{bare}$ and the averaged values of diagonal matrix elements of on-site fully screened Coulomb interaction ($\overline{W}$) at $\omega$=0.0 eV for Sn 5$p$ and S 3$p$ orbitals. All numbers are in unit of eV.}}
\resizebox{0.45\textwidth}{!}{
\begin{tabular}{@{\extracolsep{\fill}} c c c c c c c c c c c c c c c} 
\hline\hline
& & & & & SnS & & & & & \\
\hline

& & & \multicolumn{4}{c}{Bare Coulomb interaction ({$\overline{\upsilon}$})} & & &\\
\hline
                          
& \multicolumn{3}{c}{Sn 5$p$} & & & \multicolumn{3}{c}{S 3$p$} & &\\
\hline
& \multicolumn{1}{c}{diagonal} & & \multicolumn{1}{c}{off-diagonal} & & & \multicolumn{1}{c}{diagonal} & & \multicolumn{1}{c}{off-diagonal} & \\

& 4.37 & & 3.11 & & & 3.95 & & 3.04 &\\ 
 \hline
 \hline

& & \multicolumn{6}{c}{Inter-atomic Coulomb interactions} & & &\\
\hline
                               
& \multicolumn{3}{c}{Bare ({$\overline\upsilon_{Sn-S}$})} & & & \multicolumn{3}{c}{Screened ({$\overline W_{Sn-S}$})} & &\\
\hline
& \multicolumn{1}{c}{diagonal} & & \multicolumn{1}{c}{off-diagonal} & & & \multicolumn{1}{c}{diagonal} & & \multicolumn{1}{c}{off-diagonal} & \\

& 2.86 & & 2.78 & & & 0.78 & & 0.71 &\\

\hline
\hline 
& & \multicolumn{6}{c}{On-site} & & &\\
\hline\\

& \multicolumn{3}{c}{$J_{bare}$} & & & \multicolumn{3}{c}{$\overline{W}$ at $\omega$=0.0 eV} & &\\
\hline
& \multicolumn{1}{c}{Sn 5$p$} & & \multicolumn{1}{c}{S 3$p$} & & & \multicolumn{1}{c}{Sn 5$p$} & & \multicolumn{1}{c}{S 3$p$} & \\
& 0.041 & & 0.035 & & & 2.15 & & 1.8 & \\

\hline\hline

\end{tabular}}
\end{table} 

\begin{table}
\caption{\small{Averaged values of diagonal and off-diagonal matrix elements of bare Coulomb interaction ({$\overline{\upsilon}$}) for Sn (Se) 5$p$ (4$p$) orbitals \& matrix elements of bare ({$\overline\upsilon_{Sn-Se}$}) and screened ({$\overline W_{Sn-Se}$}) inter-atomic Coulomb interactions. The on-site bare exchange interaction ($J_{bare}$) and the averaged values of diagonal matrix elements of on-site fully screened Coulomb interaction ($\overline{W}$) at $\omega$=0.0 eV for Sn 5$p$ and Se 4$p$ orbitals. All numbers are in unit of eV.}}
\resizebox{0.45\textwidth}{!}{%
\begin{tabular}{@{\extracolsep{\fill}} c c c c c c c c c c c c c c c} 
\hline\hline
& & & & & SnSe & & & & & \\
\hline
& & & \multicolumn{4}{c}{Bare Coulomb interaction ({$\overline{\upsilon}$})} & & &\\
\hline
                          
& \multicolumn{3}{c}{Sn 5$p$} & & & \multicolumn{3}{c}{Se 4$p$} & &\\
\hline
& \multicolumn{1}{c}{diagonal} & & \multicolumn{1}{c}{off-diagonal} & & & \multicolumn{1}{c}{diagonal} & & \multicolumn{1}{c}{off-diagonal} & \\

& 8.64 & & 7.89 & & & 11.01 & & 9.84 &\\  
 \hline
 \hline

& & \multicolumn{6}{c}{Inter-atomic Coulomb interactions} & & &\\
\hline
                               
& \multicolumn{3}{c}{Bare ({$\overline\upsilon_{Sn-Se}$})} & & & \multicolumn{3}{c}{Screened ({$\overline W_{Sn-Se}$})} & &\\
\hline
& \multicolumn{1}{c}{diagonal} & & \multicolumn{1}{c}{off-diagonal} & & & \multicolumn{1}{c}{diagonal} & & \multicolumn{1}{c}{off-diagonal} & \\

& 2.83 & & 2.85 & & & 0.14 & & 0.15 &\\
\hline
\hline
& & \multicolumn{6}{c}{On-site} & & &\\
\hline\\
        
&  \multicolumn{3}{c}{$J_{bare}$} & & & \multicolumn{3}{c}{$\overline{W}$ at $\omega$=0.0 eV} & &\\
\hline
& \multicolumn{1}{c}{Sn 5$p$} & & \multicolumn{1}{c}{Se 4$p$} & & & \multicolumn{1}{c}{Sn 5$p$} & & \multicolumn{1}{c}{Se 4$p$} & \\
& 0.59 & & 0.62 & & & 1.6 & & 2.1 & \\

\hline\hline
 
\end{tabular}}
\end{table} 

  In present work for computing different Coulomb interactions, MLWF are formed using Sn 5$p$ and S 3$p$ (Se 4$p$) orbitals with utilizing the energy bands which are found from -8.5 eV (-6.0 eV) to 8.0 eV (7.0 eV) for SnS (SnSe). The notations used for different Coulomb interactions are explained in Section - II B. The averaged values of diagonal and off-diagonal matrix elements of on-site bare Coulomb interaction ($\overline{\upsilon}$), bare ({$\overline\upsilon_{Sn-X}$, X = S, Se) inter-atomic Coulomb interactions and screened ($\overline W_{Sn-X} $, X = S, Se) inter-atomic Coulomb interactions are tabulated in Tables. I and II for SnS and SnSe, respectively. The on-site bare exchange interaction $J_{bare}$ and the averaged values of diagonal matrix elements of on-site fully screened Coulomb interaction ($\overline{W}$) at $\omega$=0.0 eV for Sn 5$p$ and S 3$p$ (Se 4$p$) orbitals of SnS (SnSe) are also provided in Table I (Table II). The values of $\overline{\upsilon}$ (on-site $J_{bare}$) for two studied orbitals of SnSe are showing larger values than the two orbitals of SnS. The Coulomb interaction between the nuclei and electrons rises the strong attractive force when the atomic number of chalcogenide increases in Tin chalcogenides. Therefore in such case, the spread of the corresponding MLWF reduces due to the increment of localization of electrons to the respective ions. This behaviour may be responsible for increasing the values of $\overline{\upsilon}$ from SnS to SnSe. In case of inter-atomic Coulomb interactions, the bare values are almost same for both compounds, whereas the screened values of SnS are much higher than SnSe. It is known that the screened Coulomb interaction decreases faster with increasing the distance because the distance dependent screening parameter reduces the Coulomb interaction. In present scenario, when one moves from SnS to SnSe then the inter-atomic distance rises, which provides the lower value of inter-atomic Coulomb interaction for SnSe than SnS. Moreover for case of on-site, Sn 5$p$ orbitals have higher (lower) values of $\overline{W}$ at $\omega=0.0$ than S 3$p$ (Se 4$p$) orbitals for SnS (SnSe). Since, S 3$p$ orbitals are more localized than Se 4$p$ orbitals, thus Sn 5$p$ orbitals are less screened in SnS than SnSe. Now, in order to understand the correlation strength within the sample, it will be interesting to study the $\omega$ dependent Coulomb interactions for both materials.     

\begin{figure}
  \begin{center}
    \includegraphics[width=0.85\linewidth, height=10.0cm]{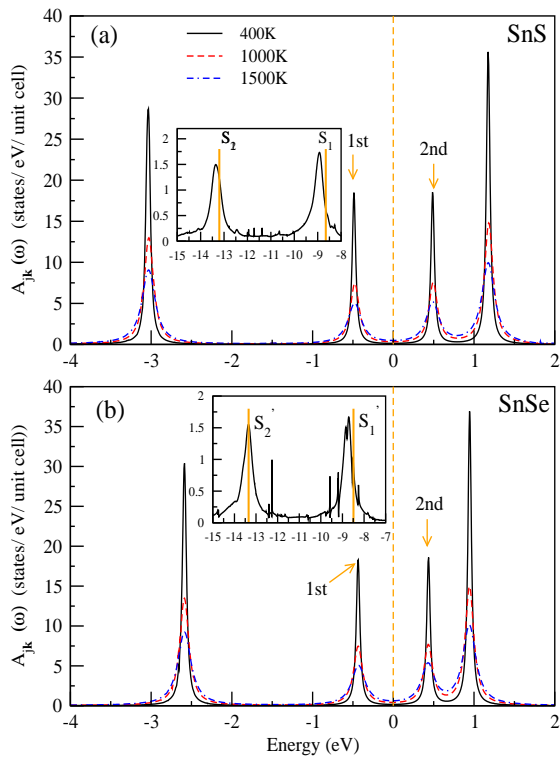} 
    \caption{(Colour online) Spectral functions of L-point from full-$GW$ at temperature 400 K (black solid line), 1000 K (red dashed line) and 1500 K (blue dotted dash line) for (a) SnS and (b) SnSe.}
    \label{fig:}
  \end{center}
\end{figure} 

\begin{figure}
  \begin{center}
    \includegraphics[width=0.8\linewidth, height=8.5cm]{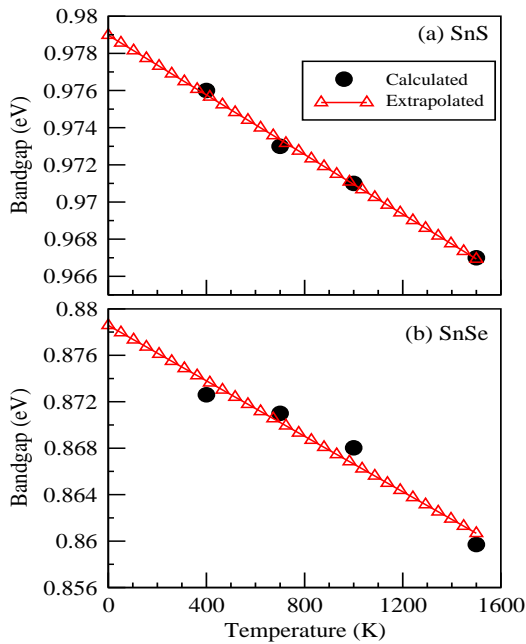} 
    \caption{(Colour online) Calculated (black dot) and extrapolated (red sold line) temperature dependent bandgap at L-point for (a) SnS and (b) SnSe.}
    \label{fig:}
  \end{center}
\end{figure}
 
  The $\omega$ dependent $\overline{W}$ and $\overline W_{Sn-X} $ (X = S, Se) have provided the information of orbital screening effect on the electronic structure of SnS and SnSe compounds. Fig. 2(a) (2(b)) shows the $\overline{W} (\omega)$ for Sn 5$p$ and S 3$p$ (Se 4$p$) orbitals along with $\overline W_{Sn-S} $ ($\overline W_{Sn-Se} $) for SnS (SnSe). The values of $\overline{W}$ of Sn 5$p$ (Se 4$p$) orbitals are slightly higher than the S 3$p$ (Sn 5$p$) orbitals for SnS (SnSe) compound. Fig. 2(a) (2(b)) is suggesting the presence of plasmon excitation of energy $\sim$14.3 eV due to the minimum value of Sn 5$p$ and S 3$p$ (Se 4$p$) orbitals of $\sim$-0.16 ($\sim$0.015) eV and $\sim$-0.47 ($\sim$-0.73) eV for SnS (SnSe) material. The presence of negative values in different form of $\omega$ dependent Coulomb interactions is usually seen when one uses the plasmon-pole approximation for computing these quantities \cite{hybertsen}. The almost similar $\omega$ dependent behaviour of $\overline W_{Sn-X} $ (X = S, Se) is found from the Figs. 2(a) and (b). The difference between the values of $\overline W_{Sn-Se} $ and Se 4$p$ orbitals are showing much larger than the difference between the values of $\overline W_{Sn-S} $ and S 3$p$ orbitals for the studied $\omega$ window. Moreover, it is noted that the strength of correlation within the sample could be possibly estimated from the oscillatory behaviour of $\omega$ dependent Coulomb interactions. Typically, it is known that highly (weakly) oscillating values of $\overline{W}(\omega)$ are observed for strongly (weakly) correlated materials with changing $\omega$ values \cite{huang2012, sakuma2012, arya2004, miyake2008}. In present case, weakly $\omega$ dependent behaviour is seen from $\overline{W}(\omega)$ curves for all studied orbitals of both materials. Therefore, the existence of weak correlation within these two samples are evident from the above discussion. In light of this to study the many-body effects of SnS and SnSe materials, the full-$GW$ method is expected to provide the appropriate descriptions of excited electronic states.       

\subsection{Finite temperature electronic structure} 

  It is known that when one calculates the electronic structure of any material using the many-body perturbation theory then it provides more closer predictive result to experimental observation as compared to single particle theory. In this scenario, the temperature dependent Green function based full-$GW$ methodology is one of the advanced method to tackle the many-body problem. Here, we have performed full-$GW$ calculation at different temperatures for SnS and SnSe. The spectral function ($A_{j\textbf{k}}(\omega)$) is defined as \cite{martin},  
 
\begin{equation}
A_{j\textbf{k}}(\omega)=-\frac{1}{\pi}Im\,[G_{j\textbf{k}}(\omega)]
\end{equation}

\begin{figure}
  \begin{center}
    \includegraphics[width=0.85\linewidth, height=9.5cm]{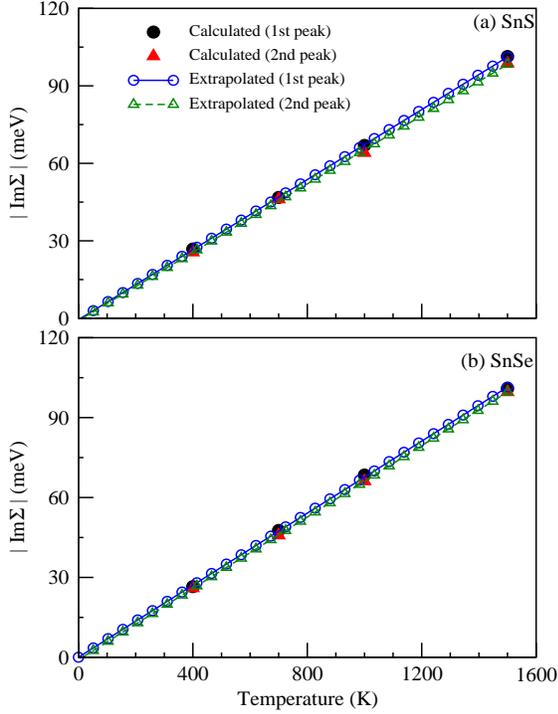} 
    \caption{(Colour online) Calculated (symbol) and extrapolated (line with symbol) $\arrowvert Im\Sigma (\omega)\arrowvert $ as function of temperatures for 1st and 2nd peak of (a) SnS \& (b) SnSe.}
    \label{fig:}
  \end{center}
\end{figure}
 
 where, $j$ denotes band index at point $\textbf{k}$ and $G_{j\textbf{k}}(\omega)$ is the Green function of many-body system. The calculated $A_{j\textbf{k}}(\omega)$ at L-point for temperatures 400 K, 1000 K and 1500 K are shown in Figs. 3(a) and (b) for SnS and SnSe, respectively. The band inversion, which is usually responsible for the presence of topological behaviour within the sample, are seen around the high-symmetric L-point for both compounds. Therefore in present case, the L-point is chosen for carrying out the detailed study of $A_{j\textbf{k}}(\omega)$ at different temperatures. The peaks' height (broadening) of $A_{j\textbf{k}}(\omega)$ is decreasing (increasing) with rise in temperature from 400-1500 K. In this case, the $E_F$ is set at the middle of 1st peak's and 2nd peak's center, which are marked in figures. The Figs. 3(a) and (b) illustrate that if one moves from SnS to SnSe then the peaks' center of VB (CB) shifted from lower (higher) to higher (lower) energy with respect to $E_F$. It is known that the atomic energy of S 3$p$ states is lower than the Se 4$p$ states. Therefore in case of VB, the resultant quasiparticle energy due to overlap of orbitals between Sn 5$p$ and S 3$p$ is lower than the overlap between Sn 5$p$ and Se 4$p$ states. The similar argument is also applicable for CB, where opposite behaviour is seen. Moreover, the hybridization energy of Sn 5$p$ - S 3$p$ seems to be stronger than Sn 5$p$ - Se 4$p$ because the value of bandgap decreases from SnS to SnSe at this k-point. Here, the energy difference between the center of 1st and 2nd peaks is considered as calculated bandgap at particular temperature. The calculated bandgaps at different temperatures are plotted in Figs. 4(a) and (b) of SnS and SnSe, where monotonically decreasing behaviour with increasing temperature is seen for both compounds due to EEI. The estimated values of bandgaps for SnS (SnSe) are found to be $\sim$0.976 ($\sim$0.873) eV, $\sim$0.973 ($\sim$0.87) eV, $\sim$0.971 ($\sim$0.868) eV and $\sim$0.967 ($\sim$0.859) eV at 400 K, 700 K, 1000 K and 1500 K, respectively. The bandgaps of SnS for the studied temperature range are showing higher values than SnSe. This behaviour suggests that the bonding and anti-bonding of Sn 5$p$ - S 3$p$ orbitals are stronger than Sn 5$p$ - Se 4$p$ orbitals at L-point. Moreover, performing the full-$GW$ calculation at lower temperature is not possible due to huge computational cost of this methodology and the limitation of present computational resources. Therefore, to compare the calculated bandgap of full-$GW$ method with the estimated value of ground state electronic structure calculations, we extrapolated the values of bandgap till 0 K with linear fitting of these calculated data. The extrapolated value of direct bandgap at 0 K is found to be $\sim$0.979 ($\sim$0.879) eV for SnS (SnSe). The difference between these values with the calculated values of $G_0W_0$ at L-point are estimated to be $\sim$0.1 eV for both materials. $G_0W_0$ provides larger bandgap value than extrapolated values at L-point for 0 K. The estimated value of bandgap at 300 K from the extrapolated data is $\sim$0.977 ($\sim$0.875) eV for SnS (SnSe). All full-$GW$ calculations are performed without including SOC because full-$GW$ calculation along with SOC is highly expensive for lower temperature. Moreover, the insight of many-body effect can be explored by studying the self-energy ($\Sigma$(\textbf{k},$\omega$)) of these two compounds. 

\begin{figure*}[]
   \begin{subfigure}{0.45\linewidth}
   \includegraphics[width=0.92\linewidth, height=7.4cm]{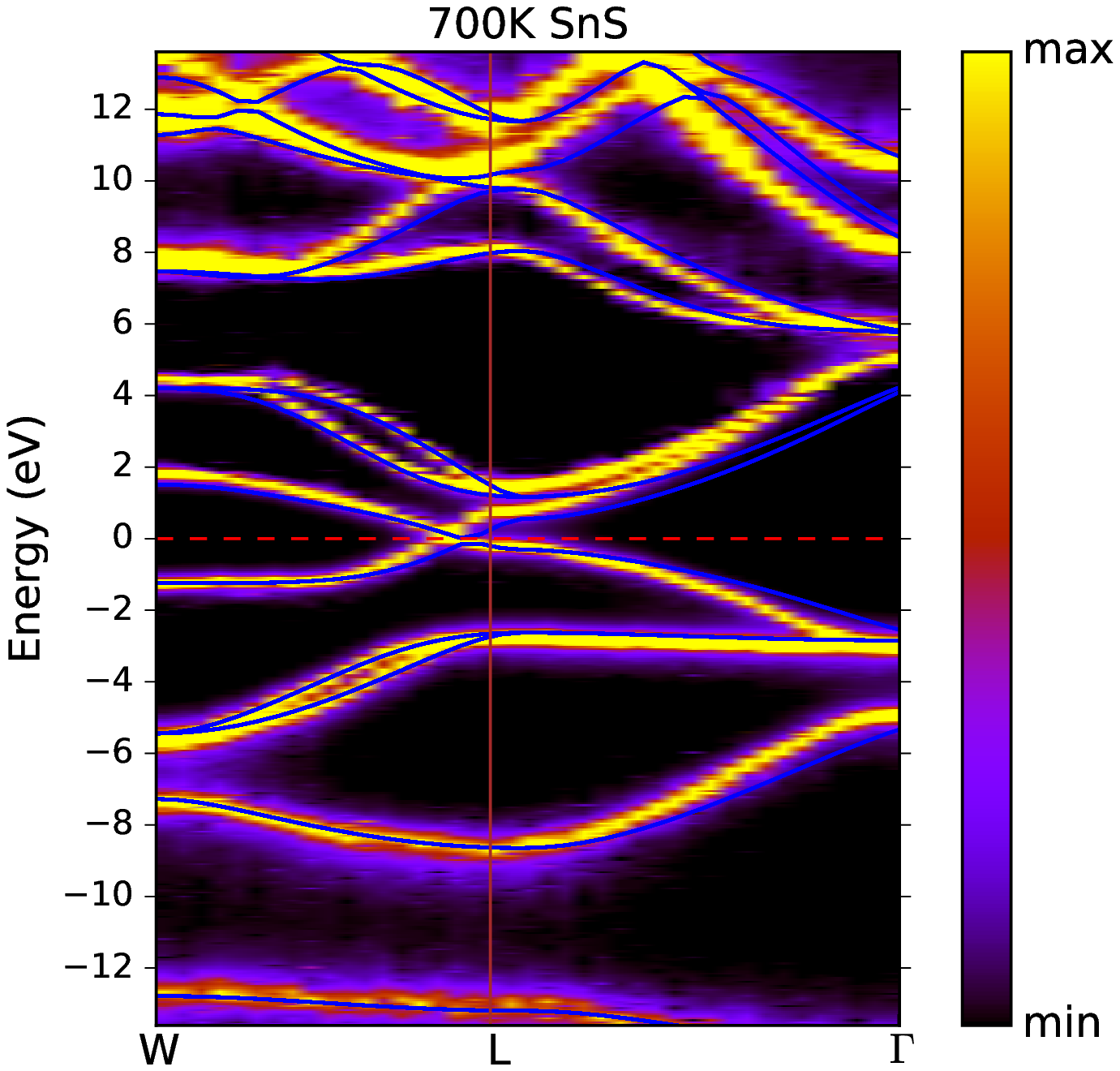}
   \caption{}
   \label{fig:} 
\end{subfigure}
\begin{subfigure}{0.45\linewidth}
   \includegraphics[width=0.92\linewidth, height=7.4cm]{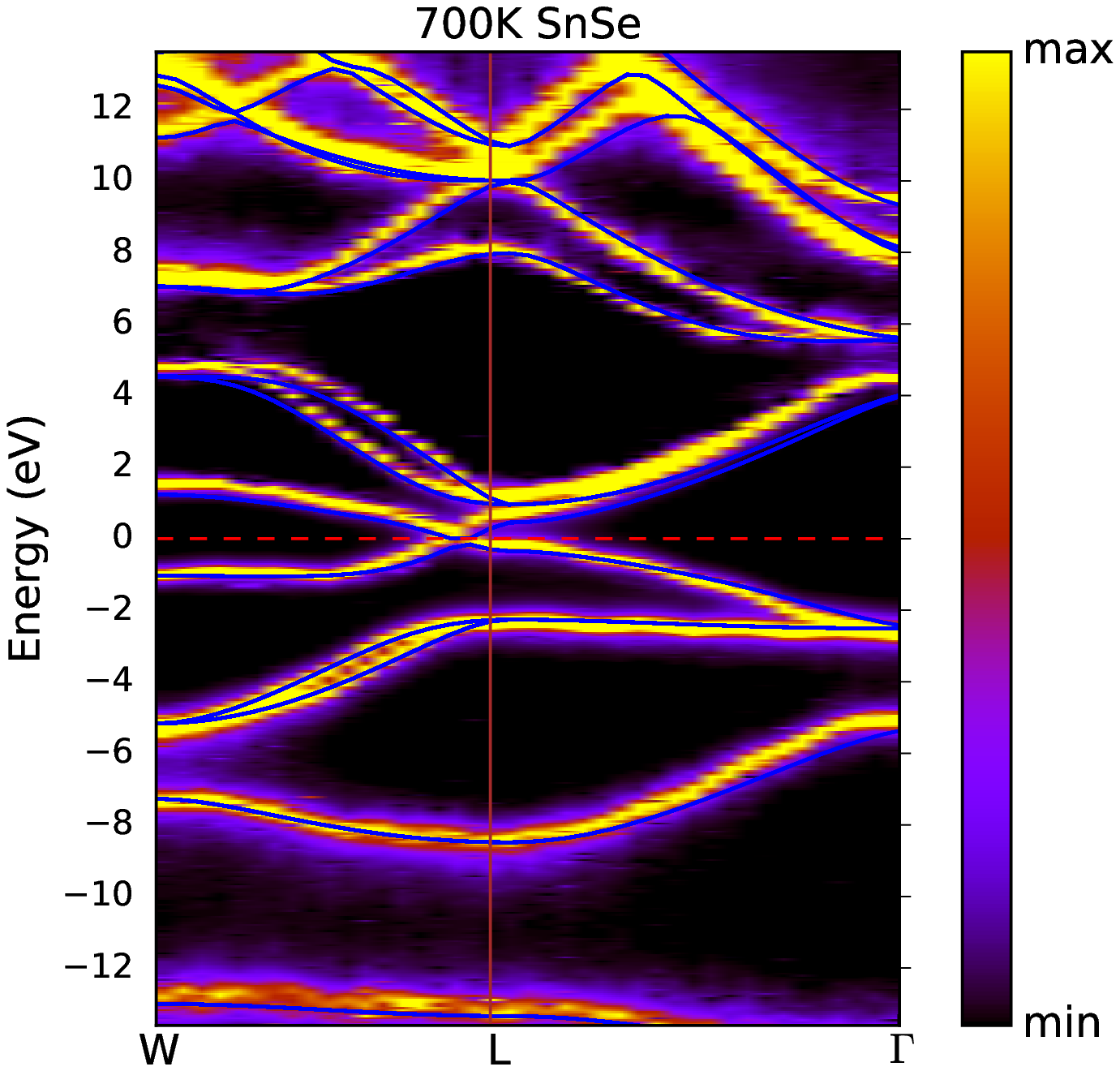}
   \caption{}
   \label{fig:}
\end{subfigure}
\caption{Momentum-resolved spectral function at 700 K of (a) SnS and (b) SnSe along W-L-$\Gamma$ direction. Blue solid lines denote the corresponding DFT bands.}
\end{figure*}
 
  The complex function $\Sigma$(\textbf{k},$\omega$) contains the detailed information about EEI of any material. The presence of spectral weight transfer is a clear evidence of many-body effect. The real part of $\Sigma$(\textbf{k},$\omega$) ($Re \Sigma (\textbf{k},\omega)$) provides the information of coherent weight of the spectrum, where the relation between $Re \Sigma (\textbf{k},\omega)$ and the mass renormalization factor ($Z_\textbf{k}(\omega)$) is given by, 
 
\begin{equation}  
Z_\textbf{k}(\omega) = \Big[1-\frac{\partial \,Re \Sigma(\textbf{k},\omega)}{\partial(\omega)}\Big]^{-1}  
\end{equation}   

  The presence of incoherent weight of the spectrum can be quantified by \big(1 - $Z_\textbf{k}(\omega)$\big). In present scenario, the values of $Z_\textbf{k}(\omega)$ are found to be $\sim$0.78 ($\sim$0.75), $\sim$0.775 ($\sim$0.77) and $\sim$0.79 ($\sim$0.77) for 1st (2nd) peak of SnS at 400 K, 700 K and 1500 K, respectively. Similarly in case of SnSe, the estimated values of this quantity are $\sim$0.77 ($\sim$0.758), $\sim$0.794 ($\sim$0.76) and $\sim$0.79 ($\sim$0.78) for 1st (2nd) peak at 400 K, 700 K and 1500 K, respectively. The computed values of $Z_\textbf{k}(\omega)$ for both these compounds are showing almost same amount for all studied temperatures. The calculated values of $Z_\textbf{k}(\omega)$ at L-point for 1st (2nd) peak, which are calculated using the ground state electronic structure calculation as performed by $G_0W_0$ method, are $\sim$0.72 ($\sim$0.73) and $\sim$0.72 ($\sim$0.725) for SnS and SnSe, respectively. Therefore, it is seen that the estimated incoherent weights of SnS (SnSe) are found to be 0.218 (0.215) and 0.236 (0.234) for 1st and 2nd peak, respectively. The presence of incoherent weight is the evidence of many-body interactions, which typically generates satellite peaks in corresponding spectrum. In order to predict the energy positions of the satellite peaks for SnS, two peaks are plotted in the inset of Fig. 3(a) along with DFT peaks (orange straight line). The peak center at $\sim$-8.9 ($\sim$-13.4) eV is marked as $S_1$ ($S_2$) in the inset of Fig. 3(a). The broadening of $S_1$ and $S_2$ peaks are showing much higher values than the 1st and 2nd peaks of this compound, where it is estimated that the $S_1$ ($S_2$) peak is $\sim$8.8 ($\sim$9.7) times more broadened than 1st peak. These behaviour suggest the existence of incoherent peaks around these energy positions along with the corresponding coherent peaks. The energy difference between $S_1$ ($S_2$) and the middle of -0.7 (-3.4) eV to 1.5 (-2.7) eV is measured to be $\sim$9.3 ($\sim$9.7) eV. Therefore, it is predicted that $S_1$ and $S_2$ are formed due to the plasmon excitations within the sample, where the plasmon frequency is estimated to be $\sim$9.5 eV for SnS. In analogues with the previous description, two peaks at $\sim$-8.7 eV and $\sim$-13.7 eV, which are corresponding to SnSe, are shown in the inset of Fig. 3(b) together with the DFT peaks (orange straight line). In the inset of Fig. 3(b), the peak at $\sim$-13.3 eV is pointed by $S^\prime_2$, whereas a kink at $\sim$-8.84 eV is marked by $S^\prime_1$. The $S^\prime_2$ is estimated to be $\sim$8.6 times more broadened than the corresponding 1st peak of SnSe. Therefore, the presence of kink at $\sim$-8.84 eV and the broadened peak of $S^\prime_2$ represent the coexistence of coherent and incoherent peaks around these mentioned energy. With the similar argument of the above discussion, the energy difference is found to be $\sim$8.75 ($\sim$9.9) eV from $S^\prime_1$ ($S^\prime_2$) to the middle of energy window -0.8 (-3.0) eV \texttwelveudash \, 1.5 (-2.17) eV. The energy difference is almost same for both the cases, where the presence of satellite peaks are predicted. Therefore, the averaged value of $\sim$9.3 eV is estimated as the plasmon frequency for SnSe. It is evident from this study that the value of plasmon frequency reduces from SnS to SnSe, which may be due to the increment of atomic radii of S to Se and the decrement of localization of S 3$p$ to Se 4$p$ orbitals. These estimated values of plasmon frequency for both materials are in good agreement with the predicted values from corresponding $\omega$ dependent Coulomb interactions plots.   
  
  Experimental studies of different transport properties suggest to carry out detailed analysis of lifetime ($\tau$) of any material for better understanding of many-body interaction within the sample. Topological non-trivial materials can be possibly distinguished from topologically trivial materials with the help of experimental evidence of transport properties. It is also noted that imaginary part of the self-energy ($Im \Sigma (\textbf{k},\omega)$) is inversely related with $\tau$. Therefore, the temperature dependent $\arrowvert Im\Sigma (\omega)\arrowvert $ of SnS and SnSe at L-point are plotted in Figs. 5(a) and (b) for further discussion on EEI. Here, the values of $\arrowvert Im\Sigma (\omega)\arrowvert $ for 1st and 2nd peaks of both materials are only focused. The monotonically increasing value of $\arrowvert Im\Sigma (\omega)\arrowvert $ with rise in temperature is observed from the figures for both materials. Therefore, the decreasing trends in $\tau$ is expected with increasing temperature for SnS and SnSe due to EEI. The calculated values of $\arrowvert Im\Sigma (\omega)\arrowvert $ are found to be $\sim$26.8 ($\sim$26.5) meV, $\sim$46.8 ($\sim$47.6) meV, $\sim$66.9 ($\sim$68.4) meV and $\sim$101.4 ($\sim$101.0) meV for 1st peak of SnS (SnSe), whereas for the 2nd peak the values are $\sim$25.6 ($\sim$26.0) meV, $\sim$46.1 ($\sim$45.8) meV, $\sim$64.1 ($\sim$66.1) meV and $\sim$98.8 ($\sim$99.6) meV at 400 K, 700 K, 1000 K and 1500 K, respectively. It is seen that the values of $\arrowvert Im\Sigma (\omega)\arrowvert $ are almost same for both SnS and SnSe for particular temperature. Therefore, it is expected to observe similar amount of $\tau$ value for SnS and SnSe due to both the peaks. Moreover, it is known that the value of $\tau$ decreases if the quasiparticle energy increases. The availability of phase space for scattering increases with rise in quasiparticle energy, which is one of the possible reason for decreasing the value of $\tau$. In present scenario, the same value of $\arrowvert Im\Sigma (\omega)\arrowvert $ at certain temperature indicates the presence of almost equal phase space for scattering due to EEI. This behaviour may be seen for these two compounds because the electronegativity of S and Se are almost same in Pauling's scale, which are 2.58 and 2.55 for S and Se, respectively \cite{electro}. Therefore, it reflects that the number of quasiparticle states (coherent) and incoherent states are almost identical for SnS and SnSe to perform the electron-electron scattering. Moreover, it is noted that such high values of $\arrowvert Im\Sigma (\omega)\arrowvert $  of both materials for the studied temperatures range suggest the importance of EEI on the transport behaviour for high temperature. In order to visualize the presence of coherent and incoherent states of both compounds, the momentum-resolved spectral function needs to be computed. 

\begin{figure}
  \begin{center}
    \includegraphics[width=0.8\linewidth, height=8.5cm]{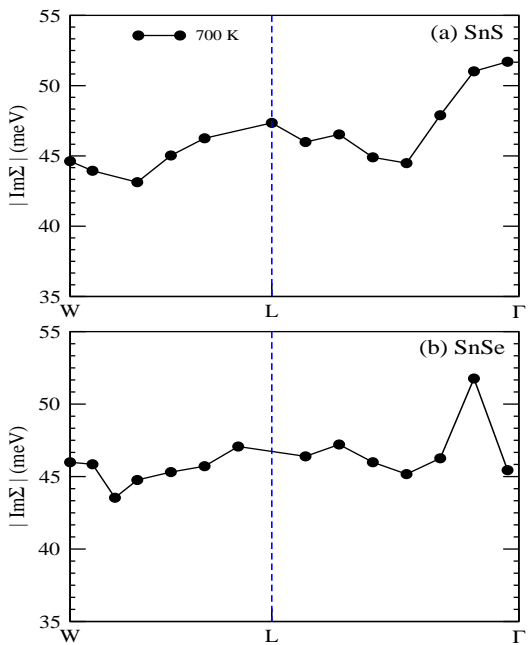} 
    \caption{(Colour online) Calculated values of $\arrowvert Im\Sigma (\omega)\arrowvert$ along W-L-$\Gamma$ direction at 700 K for the top most VB of (a) SnS and (b) SnSe.}
    \label{fig:}
  \end{center}
\end{figure}

\begin{figure}
  \begin{center}
    \includegraphics[width=0.8\linewidth, height=8.5cm]{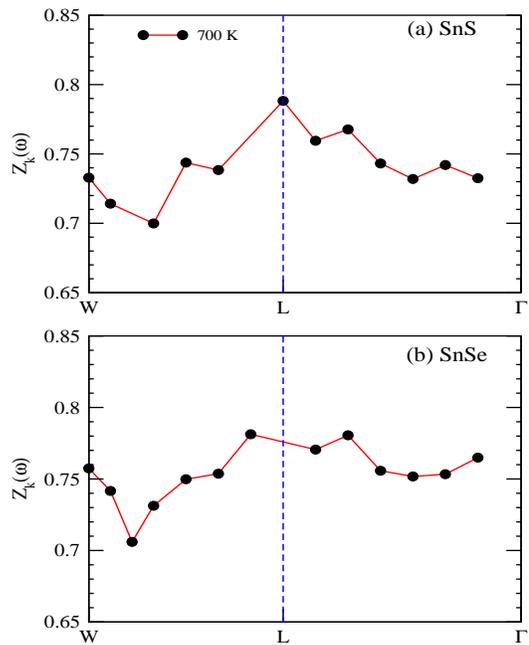} 
    \caption{(Colour online) Computed values of $Z_\textbf{k}(\omega)$ along W-L-$\Gamma$ direction at 700 K for the top most VB of (a) SnS and (b) SnSe.}
    \label{fig:}
  \end{center}
\end{figure} 
 
  Figs. 6(a) and (b) illustrate the momentum-resolved spectral function of SnS and SnSe, respectively, at temperature 700 K. These are computed using full-$GW$ method. However, performing the calculation on lower temperature is not possible due to huge computational cost. It is verified that the spectral properties of $A_{j\textbf{k}}(\omega)$ for these two compounds are not much dependent on inclusion of SOC in calculation. In these figures, the coherent (incoherent) part of the spectrum are denoted by yellowish (violetish blue) colour. In order to compare with these figures, the DFT dispersion curve of corresponding compounds are also plotted in Figs. 6(a) and (b), which are calculated using PBEsol XC functional. It is clearly seen for SnS and SnSe that the behaviour of coherent and incoherent part of the spectrum are showing almost similar nature for the W-L-$\Gamma$ direction, which is expected from both these figures. Furthermore, for more detailed explanation on EEI of SnS and SnSe, the estimated values of $\arrowvert Im\Sigma (\omega)\arrowvert $ are plotted in Figs. 7(a) and (b) for the top most VB along W-L-$\Gamma$ k-direction. The value of $\arrowvert Im\Sigma (\omega)\arrowvert $ at W-point is found to be $\sim$44.6 ($\sim$46.0) meV for SnS (SnSe). Afterwards along W-L direction, the value of $\arrowvert Im\Sigma (\omega)\arrowvert $ is first showing the decreasing nature and then increases till L-point for both these compounds. The maximum value of $\arrowvert Im\Sigma (\omega)\arrowvert $ is estimated to be $\sim$51.7 ($\sim$51.8) meV near to $\Gamma$-point for SnS (SnSe) material. These computed values are useful to compare with the experimental data of angle-resolved photoemission spectroscopy (ARPES) along this particular k-direction. To get more insight of the many-body effect through the spectral weight transfer, $Z_\textbf{k}(\omega)$ for the top most VB of SnS (SnSe) in direction of W-L-$\Gamma$ is plotted in Fig. 8(a) (8(b)). Almost similar behaviour is seen from Figs. 8(a) and (b) for $Z_\textbf{k}(\omega)$ along this direction for both compounds. The computed value of $Z_\textbf{k}(\omega)$ is found to be in range of 0.7 to 0.79 (0.71 to 0.78) for SnS (SnSe) along W-L-$\Gamma$ direction. Therefore, the range of incoherent weight for the top most VB along this mentioned k-direction is estimated to be 0.3 to 0.21 (0.29 to 0.22) for SnS (SnSe), which is usually responsible for making the satellite peaks in the corresponding spectrum.        

\section{Conclusions} 

 In present study, the famous topological crystalline insulators SnS and SnSe are focused to explore the ground and excited states using advanced $GW$ based methodology along with density functional theory. The electronic structure calculation using mBJ and $G_0W_0$ methods are shown semiconducting behaviour for both materials. But, the PBEsol provides metallic (semiconducting) ground state for SnS (SnSe) due to the improper (proper) estimation of correlation within S 3$p$ (Se 4$p$) states. The fundamental bandgaps around L-point using $G_0W_0$ (mBJ) are found to be $\sim$0.27 ($\sim$0.13) eV and $\sim$0.37 ($\sim$0.17) eV for SnS and SnSe, respectively. The behaviour of hybridization's strength between Sn 5$p$ and S 3$p$ (Se 4$p$) orbitals for SnS (SnSe) is seen to be k-point dependent. The nature of frequency ($\omega$) dependent Coulomb interactions of both materials reveal the overall strength of correlation effects within the sample and suggest full-$GW$ method for studying the excited state calculations. The electronic structure calculations of excited states for SnS and SnSe are carried out using temperature dependent full-$GW$ method in Matsubara-time domain, which have provided the linearly decreasing behaviour of temperature dependent bandgap for both materials. The presence of plasmon excitations within the spectrum is also evident from the full-$GW$ calculations for these two compounds. It is evident from both the temperature and k-point dependent imaginary part of self-energy and mass renormalization factor that the scattering mechanism due to electron-electron interaction (EEI) are found to be almost same for SnS and SnSe. All these results of many-body interactions due to EEI can be possibly verified using photoemission spectroscopic experiment.


\section{References}

\begin{thebibliography}{99}

\bibitem{hasan_rev}M. Z. Hasan and C. L. Kane, Rev. Mod. Phys. 82, 3045 (2010).
\bibitem{bansil_rev}A. Bansil, H. Lin and T. Das, Rev. Mod. Phys. 88, 021004 (2016).
\bibitem{fuprl}L. Fu, Phys. Rev. Lett. 106, 106802 (2011).
\bibitem{hsieh}T.H. Hsieh, H. Lin, J. Liu, W. Duan, A. Bansil and L. Fu, Nat. Commun. 3, 982 (2012).
\bibitem{tanaka}Y. Tanaka, Z. Ren, T. Sato, K. Nakayama, S. Souma, T. Takahashi, K. Segawa, and Y. Ando, Nat. Phys. 8, 800 (2012).
\bibitem{y_sun}Y. Sun, Z. Zhong, T. Shirakawa, C. Franchini, D. Li, Y. Li, S. Yunoki and X.-Q. Chen, Phys. Rev. B 88, 235122 (2013).
\bibitem{wang_snse_exp1}Z. Wang, J. Wang, Y. Zang, Dr. T. Jiang, Y. Gong, C.-L. Song, S.-H. Ji, L.-L. Wang, K. He, W. Duan, X. Ma, X. Chen and Q.-K. Xue, Adv. Mater. 27, 4150–4154 (2015).
\bibitem{jin_snse_exp2}W. Jin, S. Vishwanath, J. Liu, L. Kong, R. Lou, Z. Dai, J. T. Sadowski, X. Liu, H.-H. Lien, A. Chaney, Y. Han, M. Cao, J. Ma, T. Qian, S. Wang, M. Dobrowolska, J. Furdyna, D. A. Muller, K. Pohl, H. Ding, J. I. Dadap, H. G. Xing, and R. M. Osgood Jr., Phys. Rev. X 7, 041020 (2017).
\bibitem{mariano_exp_sns}A. N. Mariano and K. L. Chopra, Appl. Phys. Lett. 10, 282–284 (1967).
\bibitem{bilenkii_exp_sns2} B. F. Bilenkii, A. G. Mikolaichuk and D. M. Freik, physica status solidi (b) 28 (1), K5-K7 (1968).
\bibitem{skelton_sns}J. M. Skelton, L. A. Burton, F. Oba and A. Walsh, J. Phys. Chem. C, 121, 6446–6454 (2017).

\bibitem{wang_snse}D. Wang, W. He, C. Chang, G. Wang, J. Wang and L. D. Zhao, J. Mater. Chem. C 6, 12016–12022 (2018).
\bibitem{zhou_opto}J. Zhou, S. Zhang and J. Li, NPG Asia Materials 12, 2 (2020).
\bibitem{koller2011}D. Koller, F. Tran and P. Blaha, Phys. Rev. B 83, 195134 (2011).
\bibitem{koller2012}D. Koller, F. Tran and P. Blaha, Phys. Rev. B 85, 155109 (2012).

\bibitem{pallikara}I. Pallikara and J. M. Skelton, Phys. Chem. Chem. Phys. 23, 19219 (2021).
\bibitem{yakovkin}I. N. Yakovkin and N. V. Petrova, Phys. Lett. A 403, 127398 (2021).
\bibitem{antik}A. Sihi and S. K. Pandey, Eur. Phys. J. B 93, 9 (2020).
\bibitem{arya2006}F. Aryasetiawan, K. Karlsson, O. Jepsen and U. Sch$\ddot{o}$nberger, Phys. Rev. B 74, 125106 (2006).
\bibitem{miyake2009}T. Miyake, F. Aryasetiawan and M. Imada, Phys. Rev. B 80, 155134 (2009).
\bibitem{sakuma2013}R. Sakuma and F. Aryasetiawan, Phys. Rev. B 87, 165118 (2013).
\bibitem{amadon2014}B. Amadon, T. Applencourt and F. Bruneval, Phys. Rev. B 89, 125110 (2014).

\bibitem{vaugier}L. Vaugier, H. Jiang and S. Biermann, Phys. Rev. B 86, 165105 (2012).
\bibitem{martin}R.M. Martin, L. Reining and D.M. Ceperley, Interacting Electrons Theory and Computational Approaches (Cambridge University Press, Cambridge, 2016).
\bibitem{landau}E. M. Lifshitz and L. P. Pitaevskii, Statistical Physics, Part 2: Theory of the Condensed State. Vol. 9 (1st ed.) (1980).

\bibitem{kotlier}G. Kotliar, S.Y. Savrasov, K. Haule, V.S. Oudovenko, O. Parcollet, C.A. Marianetti, Rev. Mod. Phys. 78, 865 (2006).
\bibitem{pdjpcm}P. Dutta and S.K. Pandey, J. Phys.: Condens. Matter 31, 145602 (2019).
\bibitem{pdepl}P. Dutta and S.K. Pandey, EPL 132, 37003 (2020).

\bibitem{wien2k}P. Blaha, K.Schwarz, F. Tran, R. Laskowski, G.K.H. Madsen and L.D. Marks, J. Chem. Phys. 152, 074101 (2020).
\bibitem{pbesol}J.P. Perdew, A. Ruzsinszky, G.I. Csonka, O.A. Vydrov, G.E. Scuseria, L.A. Constantin, X. Zhou and K. Burke, Phys. Rev. Lett. 100, 136406 (2008).
\bibitem{mbj}F. Tran and P. Blaha, Phys. Rev. Lett. 102, 226401 (2009).
\bibitem{burton}L. A. Burton and A. Walsh, J. Phys. Chem. C 116, 24262−24267 (2012).
\bibitem{jiang1}H. Jiang and P. Blaha, Phys. Rev. B 93, 115203 (2016).
\bibitem{jiang2}H. Jiang, R.I. Gomez-Abal, X. Li, C. Meisenbichler, C. Ambrosch-Draxl and M. Scheffler, Comput. Phys. Commun. 184, 348 (2013).
\bibitem{elk}http://elk.sourceforge.net
\bibitem{pade}H. Vidberg and J. Serene, J. Low Temp. Phys. 29, 179 (1977).

\bibitem{antik_jpcm}A. Sihi and S. K. Pandey, J. Phys.: Condens. Matter 33, 225505 (2021).
\bibitem{hybertsen}M. S. Hybertsen and S. G. Louie, Phys. Rev. B 34, 5390 (1986).

\bibitem{jiang_mbj}H. Jiang, J. Chem. Phys. 138, 134115 (2013).

\bibitem{huang2012}L. Huang and Y. Wang, EPL 99, 67003 (2012).
\bibitem{sakuma2012}R. Sakuma, T. Miyake and F. Aryasetiawan, Phys. Rev. B 86, 245126 (2012).
\bibitem{arya2004}F. Aryasetiawan, M. Imada, A. Georges, G. Kotliar, S. Biermann and A. I. Lichtenstein, Phys. Rev. B 70, 195104 (2004).
\bibitem{miyake2008}T. Miyake and F. Aryasetiawan, Phys. Rev. B 77, 085122 (2008).
\bibitem{electro}A. L. Allred, J. Inorg. Nucl. Chem. 17, 215 (1961).



\end{thebibliography}
\end{document}